\documentclass{ws-ijmpb}

\begin{document}

\markboth{K. Shizuya}
{Direct-exchange duality of the Coulomb interaction \dots}

%
\catchline{}{}{}{}{}
%

\title{Direct-exchange duality of the Coulomb\\ interaction and collective excitations\\
in graphene in a magnetic field }

\author{K. SHIZUYA}

\address{Yukawa Institute for Theoretical Physics\\
Kyoto University,~Kyoto 606-8502,~Japan\\
shizuya@yukawa.kyoto-u.ac.jp}

\maketitle

\begin{history}
\received{31 January 2017}
\accepted{27 March 2017}
Published{\ 8 June 2017}
\end{history}

\begin{abstract}
In a magnetic field two-dimensional (2d) electron systems host, 
with quenched kinetic energy, a variety of many-body correlation phenomena, 
such as interaction-driven new states and associated collective excitations over them.
In a magnetic field the two-body operators pertinent to the 2d Coulomb interaction
obey a crossing relation, with which the Coulomb interaction
is also cast into the form of manifest exchange interaction. 
It is shown that active use of this direct/exchange duality of the interaction 
allows one to develop, within the framework of the single-mode approximation,
a new efficient algorithm for handling a wide class of collective excitations. 
The utility of our algorithm is demonstrated by studying some examples of 
inter- and intra-Landau-level collective excitations in graphene and in conventional electron systems.

\end{abstract}


\section{Introduction}
Two-dimensional (2d) electron systems such as GaAs heterostructures$^{1}$
and graphene$^{2-4}$
attract great attention in both applications and fundamental physics 
for their novel and promising features that reflect the dynamics 
specific to two dimensions and enriched with many-body correlations.
In a magnetic field, in particular, the kinetic energy of electrons is quantized 
to form a tower of flat Landau levels and, along with such a large kinetic degeneracy, 
the Coulomb interaction between carriers essentially governs the physics of many-body correlations, 
such as the fractional quantum Hall (FQH) effect$^{5,6}$
and some exotic states arising from the interplay of interaction 
and internal degrees of freedom (spin, valley, layer, etc.).
Also of interest are collective excitations 
(such as spin waves and pseudospin waves) 
which such states support.

Inter-Landau-level excitations are also amenable to many-body effects.
For (conventional) 2d electrons with quadratic dispersion, 
cyclotron resonance takes place only between adjacent levels 
and is scarcely affected by the Coulomb interaction, 
as implied by Kohn's theorem.$^{7,8}$
The situation is quite different for graphene which develops 
a quasi relativistic pattern of Landau levels, 
with a variety of cyclotron resonance$^{9}$ 
within the conduction or valence band and across the two bands. 
Many-body corrections$^{10-13}$
to such intra- and inter-band resonance, 
e.g., energy shifts and renormalization effects, 
reveal the nature of the underlying Dirac-like electrons, 
and have indeed been observed$^{14-16}$ in experiment. 
Bilayer (and few-layer) graphene$^{17}$ is 
even richer in $\lq\lq$quasi-relativistic" effects,$^{18-23}$
such as orbital degeneracy in the lowest Landau level 
and its lifting$^{24,25}$  by many-body effects.

Among theoretical frameworks$^{26-32}$
to handle such many-body effects are mean-field theory, Hartree-Fock (HF) approximation, 
the single-mode approximation (SMA), etc.
In particular, the SMA, reformulated and adapted for quantum Hall systems 
by Girvin, MacDonald and Platzman,$^{30}$ 
is a general and powerful means of studying many-body effects in a magnetic field. 
The purpose of this paper is to elaborate on the SMA 
and develop a new algorithm to facilitate actual calculations.
We first note that  in a magnetic field the two-body operators 
pertinent to the Coulomb interaction obey a crossing relation, 
with which the normal direct form of interaction
is also cast into the form of manifest exchange interaction.
Active use of this direct/exchange duality of the interaction allows one 
to effectively replace the calculation of two-body correlation functions 
(the static structure factors) crucial to the SMA  
by a far simpler calculation of the expectation values of some one-body charges.
We study some examples of inter- and intra-Landau-level collective excitations 
to demonstrate the utility of the new algorithm and
to supply some relevant techniques.

In Sec.~2 we refer to the case of graphene and set up notation 
for handling general 2d electrons in a magnetic field.  
In Secs.~3 and 4 we elaborate on the framework of the SMA,  
note the direct/exchange duality of the Coulomb interaction, 
and formulate our algorithm for general inter-Landau-level excitations. 
In Sec.~5 we examine many-body corrections to cyclotron resonance in graphene.
In Sec.~6 we extend our algorithm to intra-Landau-level collective excitations 
in flavor (spin, valley, etc) space.
In Sec.~7 we study how to handle genuine density fluctuations in our approach,
such as those over the FQH states.
Section 8 is devoted to a summary and discussion.

\section{Electrons in a magnetic field} 
The electrons in graphene are described by two-component spinors 
on two inequivalent lattice sites $(A,B)$. 
They acquire a linear spectrum (with velocity $v \sim 10^{6}$m/s) 
near the two inequivalent Fermi points $(K,K')$ in momentum space, 
and are described by an effective Hamiltonian of the form,$^{33}$
\begin{eqnarray} 
H &=&\int dx dy \, \{ \psi^{\dag} {\cal H}_{+}\psi + \chi^{\dag} {\cal H}_{-}\chi \},  \nonumber\\
{\cal H}_{\pm} &=&  v\, (\Pi_{1}\sigma^{1}+ \Pi_{2}\sigma^{2} \pm \delta m\, \sigma^{3} ) 
-eA_{0},
\label{H_GR}
\end{eqnarray}
where $\Pi_{i}= p_{i}+eA_{i}$  [with $(i=(1,2)$ or $(x,y)$] 
involve coupling to external potentials 
$(A_{i}, A_{0})$ and $\sigma^{i}$ denote Pauli matrices.
The Hamiltonians ${\cal H}_{\pm}$ describe electrons 
at two different valleys $a \in (K,K')$,
and $\delta m$ stands for a possible tiny sublattice asymmetry; 
we take $\delta m > 0$, without loss of generality.

Let us place graphene in a uniform magnetic field $B_{z}=B>0$ 
by setting $A_{i}=(-By,0)$.
The electron spectrum then forms an infinite tower of 
Landau levels of energy 
\begin{equation} 
\epsilon_{n} =s_{n}\,  \omega_{c} \sqrt{|n|+\mu^{2}}
\end{equation}
at each valley (with $s_{n}\equiv  {\rm sgn}[n] = \pm1$), 
labeled by integers $n \in (0,\pm 1, \pm2, \dots)$ and
$p_{x}$, of which only the $n=0$ (zero-mode) levels split in valley
(hence to be denoted as $n=0_{\pm}$), 
\begin{equation}
\epsilon_{0_{\mp}}= \mp  v\, \delta m = \mp \omega_{c}\,  \mu   \ \ {\rm for}\ K/K'.
\end{equation}
Here we have set, along with magnetic length $\ell \equiv 1/\sqrt{eB}$, 
\begin{equation}
\omega_{c}\equiv \sqrt{2}\, v/\ell 
\approx 36.3 \times v[10^{6}{\rm m/s}]\, \sqrt{B[{\rm T}]}\ {\rm meV},
\ \ \mu\equiv  \ell\, \delta m/\sqrt{2}.
\end{equation}

The eigenmodes at each valley $a$ are written as 
\begin{equation}
\psi_{n}^{a} = \big( | |n|-1 \rangle\, b_{n}^{a}, | |n| \rangle\, c_{n}^{a} \big)^{\rm t}
\label{psi_n}
\end{equation}
[here only the orbital eigenmodes are shown using the harmonic-oscillator basis 
$\{ |n\rangle\}$],
with $(b_{n}, c_{n})$ given by
\begin{eqnarray} 
(b_{n}^{K}, c_{n}^{K}) &=& \textstyle{1\over{\sqrt{2}}}\, (\alpha_{n}^{+}, -s_{n} \alpha_{n}^{-})
\stackrel{n\not=0, \mu \rightarrow 0}{\rightarrow}  {1\over{\sqrt{2}}}\, (1, -s_{n} ), \nonumber\\
(b_{0_{-}}^{K}, c_{0_{-}}^{K}) &=& (0, 1),    
\end{eqnarray}
where 
$\alpha_{n}^{\pm }= \sqrt{1\pm s_{n}\delta_{n} }$ 
and $\delta_{n}=  \mu/\sqrt{\mu^2 + |n|}$;
$\delta_{n} \ll 1$ for $n\not=0$ while $\delta_{0_{\pm}} =1$.

One can pass to another valley $K'$ by noting the relation 
$\sigma^{3}\, {\cal H}_{-}\sigma^{3} = -{\cal H}_{+}$.
This means that the two valleys are related as
\begin{equation}
\epsilon_{n}^{K'} = - \epsilon_{-n}^{K},\  
(b_{n}^{K'},  c_{n}^{K'})=(b_{-n}^{K}, -c_{-n}^{K}).
\label{KandKprime}
\end{equation}
Thus the Landau-level spectra as a whole are electron-hole symmetric. 
For $\delta m \rightarrow 0$ 
the two valleys differ only by the $n=0_{\pm}$ modes,
with  $(b_{0_{\pm}}, c_{0_{\pm}}) = (0, \mp1)$,
$(b_{n}, c_{n})\stackrel{n\not=0}{\rightarrow}  {1\over{\sqrt{2}}}\, (1, -s_{n} )$
and $\epsilon_{n} \rightarrow s_{n}\,  \omega_{c} \sqrt{|n|}$.

The Landau-level structure is made explicit by passing to
the $|n,y_{0}\rangle$ basis  (with $y_{0}\equiv \ell^{2}p_{x}$) 
via the expansion
$(\psi, \chi) 
= \sum_{n, y_{0}} \langle {\bf x}| n, y_{0}\rangle\, \{ \psi^{n;a}_{\alpha}(y_{0})\}$, 
where $n$ refers to the Landau level, 
$a \in (K,K')$ to the valley and
$\alpha \in (\downarrow, \uparrow)$ 
to the spin.
The Lagrangian thereby reads 
\begin{equation}
L= \int dy_{0} \sum_{n}\sum_{a,\alpha}
(\psi^{n;a}_{\alpha})^{\dag}(i\partial_{t} -\epsilon_{n}^{a}) \psi^{n;a}_{\alpha}
\label{Lag_onebody}
\end{equation}
and the charge density 
$\rho_{-{\bf p}} =\int d^{2}{\bf x}\,  e^{i {\bf p\cdot x}}\,\rho$ 
with $\rho = \psi^{\dag} \psi +  \chi^{\dag} \chi$ 
is written as$^{12}$
\begin{eqnarray}
\rho_{-{\bf p}} &=& \gamma_{\bf p}\sum_{m, n =-\infty}^{\infty}
\sum_{a,\alpha} g^{m n;a}_{\bf p}\, 
R^{m n;aa}_{\alpha\alpha; -{\bf p}}, \nonumber\\
R^{mn;ab}_{\alpha\beta; -{\bf p}}&\equiv& \int dy_{0}\,
{\psi^{m;a}_{\alpha}}^{\dag}(y_{0})\, e^{i{\bf p\cdot r}}\,
\psi^{n;b}_{\beta} (y_{0}),
\label{chargeoperator}
\end{eqnarray}
with $\gamma_{\bf p} =  e^{- \ell^{2} {\bf p}^{2}/4}$. 
Here ${\bf r} = (i\ell^{2}\partial/\partial y_{0}, y_{0})$
stands for the center coordinate with uncertainty 
$[r_{x}, r_{y}] =i\ell^{2}$.
This leads to the composition law $e^{i {\bf p\cdot r}} e^{i {\bf k\cdot r}} 
= e^{-i{1\over{2}} \ell^{2}  {\bf p \times k}}\, e^{i {\bf (p+k)\cdot r} }$,
or equivalently, 
the $W_{\infty}$ algebra$^{30}$ of the charge operators,
\begin{equation}
[ R^{jk}_{\bf k} , R^{mn}_{\bf p}] 
= \delta^{km}  \eta_{\bf k,p}\, R^{jn}_{\bf k+p}
- \delta^{nj} \eta_{\bf p,k}\, R^{mk}_{\bf k+p} ,
\label{Winf_algebra}
\end{equation}
with $\eta_{\bf k,p}\! \equiv  e^{- i {1\over{2}} \ell^2 {\bf k \times p} }$
and ${\bf k\! \times\! p} \equiv k_{x}p_{y}-k_{y}p_{x}$.
Here, for notational simplicity, we have suppressed spin and valley labels. 
Actually it is convenient to treat them collectively with the level label $n\rightarrow (n,a,\alpha)$. 
 One may regard, e.g., $R^{jk}_{\bf k}$ as $R^{jk;ab}_{\alpha\beta;{\bf k}}
\equiv R^{(j,a,\alpha), (k,b,\beta)}_{\bf k}$, and 
$\delta^{jk}$ as $\delta^{(j,a,\alpha), (k,b,\beta)} 
=\delta^{jk}  \delta^{ab} \delta^{\alpha \beta}$.
The valley and spin labels are thereby properly recovered  in Eq.~(\ref{Winf_algebra}).
Accordingly we shall often suppress them in what follows.

The coefficient matrix $ g^{m n;a}_{\bf p}$ 
at valley $a$ is given by\footnote{
We remark that an alternative choice of the U(1) phase 
of the harmonic-oscillator basis, $|n\rangle \rightarrow (e^{i \alpha})^{n} |n\rangle$,
allows one to replace $c_{n} \rightarrow e^{i\alpha} c_{n}$ in Eq.~(\ref{psi_n}) 
and $p=p_{y} + ip_{x} \rightarrow e^{i\alpha}\, p$ in $f^{kn}_{\bf p}$,
which is essentially a rotation in $xy$ plane.
} 
\begin{equation}
g^{m n;a}_{\bf p} = b_{m}^{a}\, b_{n}^{a}\, f_{\bf p}^{|m|-1,|n|-1}
+ c_{m}^{a}\, c_{n}^{a}\, f_{\bf p}^{|m|,|n|}, 
\label{gkn}
\end{equation}
where
\begin{equation}
f^{m n}_{\bf p} 
= \sqrt{n!/m!}\,
({ i \ell p/\sqrt{2}} )^{m-n}\, L^{(m-n)}_{n}
(\textstyle{1\over{2}}\ell^{2}{\bf p}^{2})
\end{equation}
for $m \ge n\ge0$, and $f^{n m}_{\bf p} = (f^{m n}_{\bf -p})^{\dag}$;
$p=p_{y}\! +i\, p_{x}$; it is understood that 
$f^{mn}_{\bf p}=0$ for $m<0$ or $n<0$.
In view of Eq.~(\ref{KandKprime}),  
$g^{m n;a}_{\bf p}$ at the two valleys are related as
\begin{equation}
g^{mn;K'}_{\bf p}= g^{-m,-n;K}_{\bf p}.
\end{equation}
Some explicit forms of $g^{m n;a}_{\bf p}$ are 
\begin{equation}
g^{00}_{\bf p} = 1, \ \
g^{11}_{\bf p} =1-(c_{1})^{2}\, \textstyle{1\over{2}}\ell^{2}{\bf p}^{2}, \
g^{10}_{\bf p} = ic_{1} \ell\, p/\sqrt{2}, \ \
g^{01}_{\bf p} = i c_{1} \ell\, p^{\dag}/\sqrt{2},
\label{example_g}
\end{equation}
with 
$c_{1}^{a} \approx - (1\mp \mu/2)/\sqrt{2}$
for $a=K/K'$.

From now on we frequently suppress
summations over levels $n$, valleys $a$ and spins $\alpha$, 
with the convention that the sum is taken over repeated indices.
The one-body Hamiltonian $H$ is thereby written as
\begin{equation}
H = \epsilon^{a}_{n}\, R^{nn;aa}_{\beta\beta;{\bf 0}} 
- \mu_{\rm Z}\, (\textstyle{1\over{2}}\sigma^{3})_{\alpha \beta} R^{nn;aa}_{\alpha\beta;{\bf 0}}. 
\label{Hzero}
\end{equation}
Here, for generality, the Zeeman term $\mu_{\rm Z} \equiv g^{*}\mu_{\rm B}B$ 
is introduced.

The Coulomb interaction 
$V= {1\over{2}} \sum_{\bf p} v_{\bf p}\,  {:\! \rho_{\bf -p}\, \rho_{\bf p}\!:}$ 
is written as
\begin{equation}
V = {1\over{2}} \sum_{\bf p}
v_{\bf p}\,\gamma_{\bf p}^{2}\,  
g^{jk; a}_{\bf p}\, g^{m n;b}_{\bf -p}
:\! R^{j k;aa}_{\alpha\alpha;{\bf -p}}\, 
R^{m n;bb}_{\beta\beta;{\bf p}}\! : , 
\label{VCoul}
\end{equation}
with the potential $v_{\bf p}= 2\pi \alpha/(\epsilon_{\rm b} |{\bf p}|)$,
$\alpha \equiv e^{2}/(4 \pi \epsilon_{0})$ and 
the substrate dielectric constant $\epsilon_{\rm b}$;
$\sum_{\bf p} \equiv \int d^{2}{\bf p}/(2\pi)^{2}$ and we set 
$\delta_{\bf p,0}\! \equiv (2\pi)^2 \delta^{2}({\bf p})$.   
As usual, normal ordering is defined as  
$:\! R^{j k} R^{m n}\! : \ \propto (\psi^{m})^{\dag} (\psi^{j})^{\dag}  \psi^{k}\psi^{n}$,
with an obvious identity 
${:R^{j k}_{\bf p} R^{m n}_{\bf q}\!:}\,  =\ :\!R^{m n}_{\bf q} R^{j k}_{\bf p}\!:$.

So far we have set up our notation for monolayer graphene 
but the total Hamiltonian of the form
$H^{\rm tot} = H + V$ with Eqs.~(\ref{Hzero}) and~(\ref{VCoul}) 
applies to general electron systems in a magnetic field as well, 
so does our analysis below.
For conventional 2d electrons, e.g.,  one may simply set 
$\epsilon_{n} \rightarrow \omega_{c} (n+ {1\over{2}})$ 
with $\omega_{c} = eB/m^{*}$
and restrict orbital labels to $n \in (0,1,2, \cdots)$ 
and  $g^{m n}_{\bf p}\rightarrow f^{m n}_{\bf p}$.

\section{Collective excitations}

Suppose now that a uniform ground state $|{\rm Gr}\rangle$ is realized 
at some filling factor in a magnetic field.
Our task is to study collective excitations over this ground state 
using the Hamiltonian $H + V$.
For definiteness, let us consider interlevel excitations 
from $\{j; a,\alpha \}$ to $\{n; b,\beta\}$, using the SMA.
The SMA is a variational method$^{30-32}$
that adopts $R^{nj; ba}_{\beta\alpha;{\bf p}}|{\rm Gr} \rangle$ 
as the trial state for such an excitation.
It is neatly systematized in the framework of effective Lagrangian.

Let $\Xi^{nj;ba}_{\beta \alpha; {\bf p}}$ be an interpolating field 
associated with the charge $R^{nj; ba}_{\beta\alpha;{\bf -p}}$ 
and denote  
\begin{equation}
\Xi R =\sum_{\bf p} \Xi^{nj}_{\bf p} R^{nj}_{\bf -p}
\end{equation}
for short;  
$(\Xi^{nj;ba}_{\beta \alpha; {\bf p}})^{\dag} =\Xi^{jn;ab}_{\alpha \beta; {\bf -p}}$
so that $\Xi R$ is hermitian; 
here we consider general $n\leftarrow j$ channels all together 
and the sum over orbital (and suppressed valley and spin) labels is understood.
One then regards interlevel excitation as a $W_{\infty}$-rotation
$e^{-i \Xi R} |{\rm Gr} \rangle$ of $|{\rm Gr} \rangle$ in the orbital space,
and evaluates the associated energy.
Note first that, via ${\cal U} = e^{i \Xi R}$, the field $\psi^{m}$ turns into
\begin{equation}
{\cal U}\, \psi^{m}(y_{0})\, {\cal U}^{-1}= 
[U^{-1}]^{mn}\, \psi^{n}(y_{0}) \equiv  \psi'^{m}(y_{0}),
\end{equation}
where $U=e^{ i\Xi [{\bf r}]}$ and $(\Xi [{\bf r}])^{mn} 
\equiv \sum_{\bf p} \Xi^{mn}_{\bf p}\, e^{i {\bf p\cdot r}}$.
Replacing $\psi$ by $\psi' = U^{-1}\psi$ in the Lagrangian~(\ref{Lag_onebody}) 
and taking the expectation value $\langle {\rm Gr}| \cdots |{\rm Gr}\rangle$ 
then reveals the associated energy change  
in the form of Lagrangian$^{34}$ for $\Xi$,  
\begin{equation}
L_{\Xi}=\langle {\rm Gr} |{\cal U}
(i \partial_{t} -H^{\rm tot}) {\cal U}^{-1}|{\rm Gr} \rangle, 
\label{effLag}
\end{equation}
with $H^{\rm tot}= H + V$;
$\partial_{t}$ acts on $\Xi_{\bf p}$ in ${\cal U}^{-1} = e^{-i \Xi R}$.
[Note in this connection the relations
$\psi^{\dag}U{\cal H}U^{-1}\psi = {\cal U}(\psi^{\dag}{\cal H}\psi){\cal U}^{-1}$ 
and 
$\psi^{\dag}U i(\partial_{t}U^{-1})\psi = {\cal U} i\partial_{t}{\cal U}^{-1}$.]

Our task in this paper is to develop a general and efficient way to calculate 
the effective Lagrangian~(\ref{effLag}).
To this end one first needs  a set of ground-state expectation values, 
which we denote as
\begin{equation} 
\langle {\rm Gr}|R^{mn;ab}_{\alpha\beta; {\bf p}}|{\rm Gr}\rangle
= \bar{\rho}\, \nu_{n}^{a \alpha}\delta^{mn}\delta^{ab}\delta^{\alpha\beta}\, 
\delta_{\bf p,0}
\end{equation}
for good quantum numbers $\{n,a,\alpha\}$,
where $\nu_{n}^{a \alpha}$ denotes the filling fraction of the  $\{n,a,\alpha\}$ level
and  $\bar{\rho} \! \equiv 1/(2\pi \ell^2)$.
If, for example, there arises mixing in valley, 
one has to first rotate the field $\psi$ in valley space 
and define filling fractions only for a set of good (i.e., diagonal) valley labels.
In what follows we regard one-body labels $\{n,a,\alpha\}$ of $H$
as good quantum numbers and study many-body effects to $O(V)$.

The one-body part of $L_{\Xi}$ in Eq.~(\ref{effLag})
is solely governed by the expectation values  of 
rotated charges 
$(R^{mn}_{\bf p})^{\, \cal U} \equiv {\cal U} R^{mn}_{\bf p} {\cal U}^{-1} 
= e^{ i\, \Xi R}\, R^{mn}_{\bf p}\, e^{-i\, \Xi R}$.
It is useful to write 
$C_{\bf p}(R_{\bf -p})^{\cal U} \equiv C^{mn}_{\bf p}(R^{mn}_{\bf -p})^{\cal U}$,
with an arbitrary function $C^{mn}_{\bf p}$.
The rotated charges then read, to $O(\Xi^2)$,
\begin{eqnarray}
C_{\bf p}(R_{\bf -p})^{\cal U}
&=& C_{\bf p} R_{\bf -p} 
+ \sum_{\bf k} C^{(1)}_{\bf k,p}\, R_{\bf -k - p} 
+ \sum_{\bf q, k} C^{(2)}_{\bf q,k,p}\, R_{\bf -q-k -p} + \cdots,
 \\
C^{(1)}_{\bf k,p} &=& i  \{ \eta_{\bf k,p}\,(\Xi_{\bf k}C_{\bf p}) 
- \eta_{\bf p,k}\,(C_{\bf p} \Xi_{\bf k})  \},
\nonumber\\
C^{(2)}_{\bf q,k,p} &=&  \eta_{\bf k, p, q}\, (\Xi_{\bf k}C_{\bf p}\Xi_{\bf q})
-{\textstyle{1\over{2}}} \eta_{\bf k, q, p}\, (\Xi_{\bf k}\Xi_{\bf q}C_{\bf p}) 
- {\textstyle{1\over{2}}}\eta_{\bf p, k, q}\, (C_{\bf p}\Xi_{\bf k}\Xi_{\bf q}),
 \label{RupU}
\end{eqnarray}
with $\eta_{\bf k,p}\! \equiv  e^{- i {1\over{2}} \ell^2 {\bf k \times p} }$
and 
$\eta_{\bf q, k, p}\! \equiv e^{- i{1\over{2}} \ell^2 \{ {\bf  q \times k + (q+k) \times p} \} }$;
$(\Xi_{\bf k}C_{\bf p})$ stands for the matrix product 
 $(\Xi_{\bf k}C_{\bf p})^{jn}=\Xi_{\bf k}^{jk}C_{\bf p}^{kn}$, etc.
 In particular, the ground-state expectation values are neatly written as
\begin{eqnarray}
&&\langle (R^{jn;ab}_{\bf -p})^{\cal U} \rangle 
= \bar{\rho}\,\Big[ \nu^{a}_{j} \delta^{nj} \delta^{ba} \delta_{\bf p,0} 
+i ( \nu^{b}_{n} \! -\!  \nu^{a}_{j} )\, \Xi^{nj;ba}_{\bf -p} + \sum_{\bf  q}e^{ i {1\over{2}} \ell^2 {\bf p \times q} }\, 
\Gamma^{nj;ba}_{\bf p,q}
+\cdots \Big],  \nonumber\\
&& \Gamma^{nj;ba}_{\bf p,q}= \sum_{k, c} \{
\nu_{k}^{c} - {\textstyle{1\over{2}}} (\nu_{n}^{b} \!+\nu_{j}^{a})
\}\, \Xi^{nk;bc}_{\bf -q-p}\, \Xi^{kj;ca}_{\bf q},   
\end{eqnarray}
where  for generality we have restored valley indices.
From now on expectation values
$\langle {\rm Gr}| \cdots |{\rm Gr} \rangle$ will be simply denoted 
by $\langle \cdots \rangle$.

The one-body 
$\sum_{j,a} \epsilon_{j}^{a} \langle (R^{jj;aa}_{{\bf p}=0} )^{\cal U}\rangle$ term, 
in particular, leads to the $O(\Xi^{2})$ energy term
\begin{equation}
\bar{\rho}\, \sum_{n>j} \sum_{a,b}
 ( \epsilon_{n}^{b} -\epsilon_{j}^{a} )\, (\nu_{j}^{a} - \nu_{n}^{b} ) 
 \sum_{\bf  k}\Xi^{jn;ab}_{\bf -k}\Xi^{nj;ba}_{\bf k}, \ \ 
\end{equation}
where, in passing, we have cast
$\sum_{n,j} \epsilon_{j}^{a} ( \nu_{n}^{b} -\nu_{j}^{a})$ into the above form
by noting the symmetry 
of $\Xi^{jn;ab}_{\bf -k}\Xi^{nj;ba}_{\bf k}$ 
under $(b \leftrightarrow a, j \leftrightarrow n, {\bf k} \leftrightarrow  {\bf -k})$.
Similarly, the ${\cal U}\, i\partial_{t}\, {\cal U}^{-1}$ term 
$(\sim  i {1\over{2}}\, [\Xi R,  \dot{\Xi} R] )$ yields, up to a total derivative, 
\begin{equation}
i \bar{\rho} \sum_{\bf k}\sum_{j,n}\nu_{j}\,  
\Xi^{jn}_{\bf -k}\dot{\Xi}^{nj}_{\bf k} 
\approx i \bar{\rho} \sum_{\bf k} \sum_{n>j} (\nu_{j} - \nu_{n} )\,  
\Xi^{jn}_{\bf -k}\dot{\Xi}^{nj}_{\bf k},
\end{equation}
where $\dot{\Xi}\equiv \partial_{t}\Xi$.
They combine to constitute the one-body       part of the effective Lagrangian:
\begin{equation}
L^{\rm 1b}_{\Xi}  =  \bar{\rho}\, \sum_{a,b}\sum_{n>j} 
\sum_{\bf  k}
(\xi^{nj;ba}_{\bf k})^{\dag}  \big\{ i\partial_{t} 
- ( \epsilon_{n}^{b} -\epsilon_{j}^{a} ) \big\}\,  \xi^{nj;ba}_{\bf k},
 \label{Leffonebody}
\end{equation}
where 
we have rescaled $\Xi^{nj;ba}_{\bf k} = {\cal N}^{ba}_{nj}\, \xi^{nj;ba}_{\bf k}$
with ${\cal N}^{ba}_{nj}\equiv \sqrt{\nu_{j}^{a}  -  \nu_{n}^{b} }$;
$(\xi^{nj;ba}_{\bf k})^{\dag} = \xi^{jn;ab}_{\bf -k}$.
Clearly the field $\xi^{nj;ba}_{\bf k}$ describes 
an interlevel $n \leftarrow j$ excitation 
of energy $\epsilon_{n}^{b} -\epsilon_{j}^{a}$.
Note that $0<{\cal N}^{ba}_{nj}\le 1$
for each {\em open} $(n,b) \leftarrow (j,a)$ channel of excitation;
${\cal N}^{ba}_{nj}=0$ for inactive channels 
so that only open channels appear in Eq.~(\ref{Leffonebody}).

 To handle the Coulomb interaction one needs the knowledge of 
 the static structure functions
$\langle R^{ij}_{\bf p}R^{k\ell}_{\bf -p} \rangle$, 
to be examined in the next section.

\section{Direct/exchange duality of the Coulomb interaction}

In the $y_{0}$ basis the plane wave $e^{-i{\bf p \cdot r}}$ is a unitary matrix, 
$(e^{-i{\bf p \cdot r}})^{\dag}=e^{i{\bf p \cdot r}}$, 
with elements
\begin{equation}
\langle y_{0}| e^{-i{\bf p \cdot r}}|y'_{0}\rangle 
= \delta (y_{0} - y'_{0} + \ell^{2}p_{x})\, 
e^{-i{1\over{2}}p_{y} (y_{0} + y'_{0} )}.
\end{equation}
They obey the completeness relation
\begin{equation}
\sum_{\bf p}\,  \langle y'_{0}| e^{-i{\bf p \cdot r}}|y_{0}\rangle\, 
\langle z_{0}| e^{i{\bf p \cdot r}}|z'_{0}\rangle 
= \bar{\rho}\, \delta_{y_{0}, z_{0}}\,  \delta_{y'_{0},z'_{0}},
\end{equation}
as verified directly, 
where $\delta_{y_{0}, z_{0}} \equiv \delta (y_{0} - z_{0})$ 
and $\bar{\rho} \! \equiv 1/(2\pi \ell^2)$.
This relation allows one to invert the charge operators
$R^{mn}_{\bf -p} =\sum_{y_{0}, y'_{0}}
 \psi^{m\dag}(y_{0}) \langle y_{0}|e^{i{\bf p \cdot r}}|y'_{0}\rangle \psi^{n}(y'_{0})$
for the field products $\psi^{m\dag} \psi^{n}$,
\begin{equation}
\bar{\rho}\, \psi^{m\dag}(y_{0}) \psi^{n} (y'_{0}) 
=  \sum_{\bf p}\,  \langle y'_{0}| e^{i{\bf p \cdot r}}|y_{0}\rangle\, R^{mn}_{\bf p}.
\label{inversionformula}
\end{equation}
This inversion formula has long been known.$^{26,27}$


\begin{figure}[tbp]
\centerline{\psfig{file=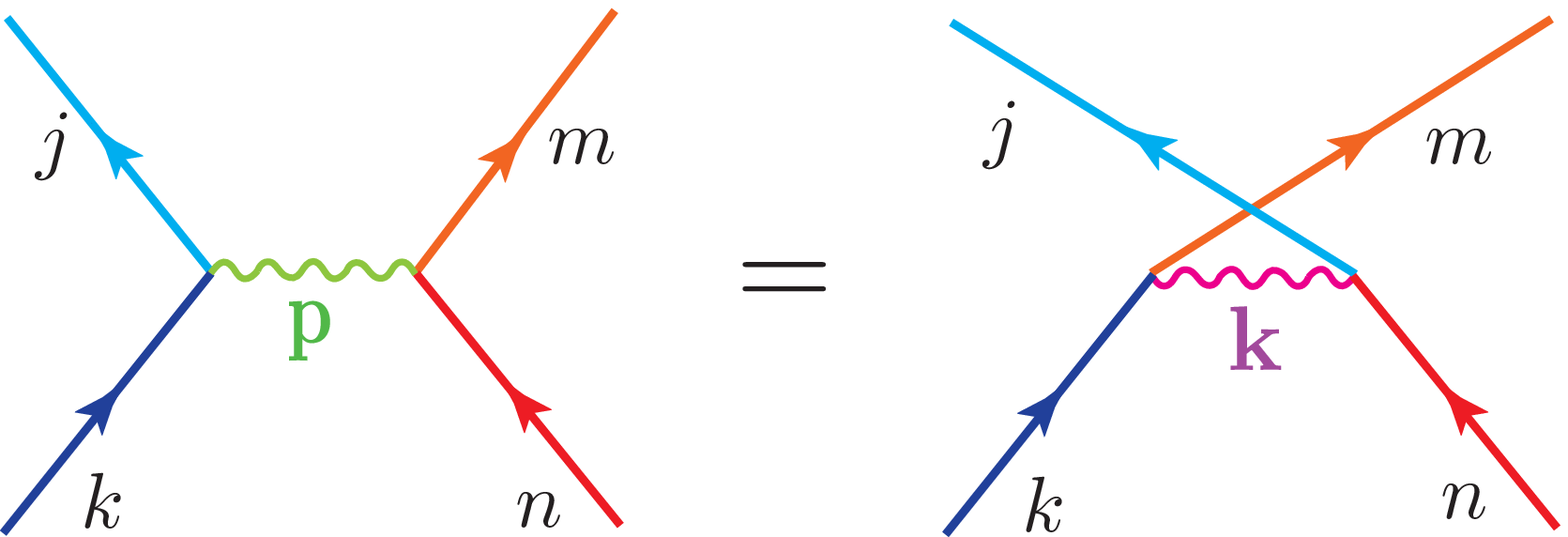,width=3.25in}}
\vspace*{8pt}
\caption{Crossing relation for $: \!R^{jk}_{\bf -p} R^{mn}_{\bf p}\!: $.
}
\end{figure}


Consider now the normal-ordered product 
${:\! R^{jk}_{\bf -p} R^{mn}_{\bf p}\!:}  
\sim \psi^{m\dag}\psi^{j\dag}
\psi^{k}\psi^{n}$ and 
let $\psi^{j\dag}$ be paired with $\psi^{n}$ 
and $\psi^{m\dag}$ with $\psi^{k}$, using Eq.~(\ref{inversionformula}).
A little algebra\footnote{
Note that 
$\int dy_{0} \langle y_{0}| e^{i {\bf p\cdot r}} |y_{0}\rangle = \bar{\rho}\, \delta_{\bf p,0}$
and $\sum_{\bf k} e^{i\, \ell^{2}  {\bf p \times k}} = (\bar{\rho})^2 \delta_{\bf p,0}$.
} 
then yields the crossing formula
\begin{equation}
: \!R^{jk}_{\bf -p} R^{mn}_{\bf p}\!: \ 
= - {1\over{\bar{\rho}}}  \sum_{\bf k}  
e^{i\, \ell^{2}  {\bf p \times k}}\,  
 :\! R^{mk}_{\bf -k} R^{jn}_{\bf k}\!:  .
\label{Xrel}
\end{equation}
See Fig.~1. For the regular product $R_{\bf -p}R_{\bf p}$ 
the crossing relation is somewhat complicated:
\begin{equation}
R^{jk}_{\bf -p} R^{mn}_{\bf p}
= - {1\over{\bar{\rho}}}  \sum_{\bf k}  
e^{i\, \ell^{2}  {\bf p \times k}}\,  
R^{mk}_{\bf -k} R^{jn}_{\bf k} + \delta^{km} R^{jn}_{\bf 0} 
+ \bar{\rho}\, \delta_{\bf p,0}\,  \delta^{kj} R^{mn}_{\bf 0} .
\end{equation}

The Coulomb interaction $V$ in Eq.~(\ref{VCoul}) has direct interaction 
and also exchange interaction at the quantum level. 
The crossing relation~(\ref{Xrel}) allows one to rewrite $V$ in the form 
of exchange interaction,
\begin{eqnarray}
&&V = -{1\over{2 \bar{\rho}}}  \sum_{\bf k}  W^{jk; mn;ab}_{\bf k} 
:R^{m k;ba}_{\beta\alpha;\bf - k}  R^{j n;ab}_{\alpha\beta;\bf k} :, \nonumber\\
&&W^{jk; mn;ab}_{\bf k} \equiv \sum_{\bf p} v_{\bf p}\gamma_{\bf p}^2\, 
 g^{j k;a}_{\bf p}\,  g^{m n ;b}_{\bf -p}\, e^{i\ell^2{\bf p \times k}}.
\label{Vdual}
\end{eqnarray}
This makes manifest,  on the operator level, the direct/exchange duality 
of the 2d Coulomb interaction  in a magnetic field.

In view of Eq.~(\ref{Xrel}) 
the normal-ordered static structure factors obey the relation
\begin{equation}
\langle :\! R^{jk}_{\bf -p} R^{mn}_{\bf p}\! : \rangle
= -{1\over{\bar{\rho}}}\sum_{\bf k}  
e^{i\, \ell^{2}  {\bf p \times k}}\,  
\langle :\! R^{mk}_{\bf -k}R^{jn}_{\bf k}\!: \rangle .
\label{DualityRel}
\end{equation}
For a general uniform many-body state $|{\rm Gr}\rangle$, 
the structure factors 
$\langle\,  {:\! R^{jk}_{\bf -p} R^{mn}_{\bf p}\!: }\,  \rangle$
are nonzero only for combinations 
$\delta^{jk}\delta^{mn}$ or $\delta^{mk}\delta^{jn}$
of labels. There are thus three cases to consider, 
(i) $\delta^{jk}\delta^{mn}$ and $j\not= m$,
(ii) $\delta^{mk}\delta^{jn}$ and $j\not= m$, and (iii) $j=k=m=n$.
For case (i), $\langle\,  {:R^{jk}_{\bf -p} R^{mn}_{\bf p}\!:} \, \rangle$ equals 
$\langle R^{jj}_{\bf -p}\rangle \langle R^{mm}_{\bf p} \rangle
= (\bar{\rho} )^2\nu_{j}\nu_{m}\, \delta_{\bf 0,0}\, \delta_{\bf p,0}$, 
i.e., proportional to
$\delta_{\bf p, 0}$, where $\delta_{\bf 0,0} = \int d^{2}{\bf x}$.
For case (ii), Eq.~(\ref{DualityRel}) implies that 
$\langle\,  {:R^{jm}_{\bf -p} R^{mj}_{\bf p}\!:} \, \rangle
= -\bar{\rho}\, \nu_{j}\nu_{m}\, \delta_{\bf 0,0}$, 
i.e., a constant independent of ${\bf p}$.

In case (iii) one encounters the static structure factor projected to the $n$th level, 
defined as 
$\langle R^{nn}_{\bf -p} R^{nn}_{\bf p} \rangle  
= \langle R^{nn}_{\bf -p}\rangle \langle R^{nn}_{\bf p} \rangle 
+ \delta_{\bf 0,0} \bar{\rho}\, \nu_{n}\, \hat{s}_{n}({\bf p})$ (for fixed $n$), or 
\begin{equation}
\langle :\! R^{nn}_{\bf -p}\, R^{nn}_{\bf p}\!: \rangle
= \delta_{\bf 0,0}\, \bar{\rho}\, \nu_{n}\, [\bar{\rho}\, \nu_{n}\, \delta_{\bf p,0} 
+\hat{s}_{n}({\bf p})  - 1];
\label{StaticSF}
\end{equation}
$\bar{\rho}\, \delta_{\bf 0,0}\, 
= \bar{\rho} \int d^{2}{\bf x}$  stands for  
the total number of electrons per filled level.
(Normally it is 
$\gamma_{\bf p}^{2}\, \hat{s}_{n}({\bf p})$ 
that is defined as the projected structure factor.$^{30}$)
Possible  ${\bf p}$ dependence 
$\hat{s}_{n}({\bf p})$ comes from nontrivial correlations 
within a partially filled level $n$,
and $\hat{s}_{n}({\bf p}) \rightarrow 0$ as $\nu_{n}\rightarrow 1$,
i.e., for a filled level.
Let us isolate a constant piece, 
$\hat{s}_{n}({\bf p}) = \hat{s}_{n}(\infty) + \delta \hat{s}_{n}({\bf p})$, 
so that  
$\delta \hat{s}_{n}({\bf p})$ has a Fourier transform.
Substituting Eq.~(\ref{StaticSF}) into Eq.~(\ref{DualityRel}) then implies that 
\begin{equation}
\hat{s}_{n}(\infty) = 1- \nu_{n},\ \ 
\delta \hat{s}_{n}({\bf p}) =
-(1/\bar{\rho})\sum_{\bf k}  \delta \hat{s}_{n}({\bf k})\, 
e^{i\, \ell^{2}  {\bf p \times k}}.
\end{equation}
Actually, $\hat{s}_{n}(\infty) = 1- \nu_{n}$ is consistent with 
the HF approximation, which leads$^{29}$
to $\hat{s}_{n}({\bf p}) = 1- \nu_{n}$.
Thus $\delta \hat{s}_{n}({\bf p})$ stands for a possible deviation 
from the HF treatment, 
and such a deviation is present in the exact result.
Indeed, for Laughlin's wave function$^{6}$ 
for the FQH states with $\nu=1/3, 1/5, \cdots$,
e.g., one knows quite generally$^{30}$ that 
$\hat{s}_{n=0}({\bf p}) \rightarrow 0$ as ${\bf p} \rightarrow 0$ 
while 
$\hat{s}_{0}({\bf p}) \rightarrow 1- \nu$ for large $|{\bf p}|$.
Thus $\delta \hat{s}_{n}({\bf p})$ arises for small $|{\bf p}|$. 
It will be enlightening here to understand the HF-approximation result 
in the following way: 
Suppose that the $n$th level 
simply consists of filled modes $\{|y_{0}\rangle\}_{\rm fill}$ of fraction $\nu_{n}$ 
and empty modes $\{|y'_{0}\rangle\}_{\rm emp}$ of fraction $1-\nu_{n}$ 
in $y_{0}$ space. 
For such a configuration the structure factor associated with 
intralevel transitions 
$\{|y_{0}\rangle\}_{\rm fill} \rightarrow \{|y'_{0}\rangle\}_{\rm emp}
\rightarrow \{|y_{0}\rangle\}_{\rm fill}$ 
is calculated in essentially the same way as in case (ii), 
yielding
$\langle {:\! R^{nn}_{\bf -p} R^{nn}_{\bf p}\! :}\rangle
\rightarrow -\bar{\rho}\, \nu_{n}\, (1- \nu_{n})\, \delta_{\bf 0,0}$.
Thus $\hat{s}_{n}({\bf p}) = 1- \nu_{n}$ is a possible exact result 
for the simple mean-field-like configuration we have supposed.

In reality it is a hard task to calculate ${\bf p}$-dependent correlation 
$\delta \hat{s}_{n}({\bf p})$ for general configurations;
one, e.g.,  has to resort to an exact diagonalization study.\footnote{
See, e.g., K. Asano and T. Ando, in Ref.~\refcite{MBE}.   
} 
Let us for the moment ignore $\delta \hat{s}_{n}({\bf p})$, 
which, if needed, is easily and formally recovered.
Then  $\langle\, {: \! R^{jk}_{\bf -p} R^{mn}_{\bf p}\!:} \, \rangle$ 
consists of the delta-function piece $\langle R^{jk}_{\bf -p}\rangle 
\langle R^{mn}_{\bf p} \rangle$ and a constant piece,
\begin{equation}
\langle\, {: \! R_{\bf -p} R_{\bf p}\!:} \, \rangle 
= (\delta_{\bf p,0}\,  {\rm piece} ) + ({\rm constant}).
\end{equation}
Substituting this form into the crossing relation~(\ref{DualityRel})
reveals that the $\delta_{\bf k,0}$ piece and constant piece of 
$\langle\, {:\! R^{mk}_{\bf -k} R^{jn}_{\bf k}\!:}\,  \rangle$ 
turn into the constant piece and $\delta_{\bf p,0}$ piece of 
$\langle\, {:\! R^{jk}_{\bf -p} R^{mn}_{\bf p}\!:} \, \rangle$, respectively.
Here we see a dual relation: the singular and constant pieces are interchanged 
between the {\em dual} pair of normal-ordered structure factors.
As a result, one can evaluate 
$\langle\, {:\! R_{\bf -p}R_{\bf p}\!:} \,\rangle$ 
via the expectation values  $\langle R_{\bf p} \rangle$ alone, 
\begin{eqnarray}
\langle :\! R^{jk}_{\bf -p} R^{mn}_{\bf p}\! : \rangle 
&=& \langle R^{jk}_{\bf -p} \rangle \langle R^{mn}_{\bf p} \rangle 
-{1\over{\bar{\rho} }}\sum_{\bf k}  e^{ i \ell^{2}  {\bf p \times k}}
\langle R^{mk}_{\bf -k} \rangle \langle R^{jn}_{\bf k} \rangle, 
\label{RRfromR}
\\
&=& \delta_{\bf 0,0}\, \bar{\rho}\, \, \nu_{j}  \nu_{m}\, 
(\delta^{jk}\delta^{mn}\, \bar{\rho}\,\delta_{\bf p,0}  
-\delta^{mk}\delta^{jn} ).
\end{eqnarray}

Let us now consider the Coulombic corrections  
$\langle V^{\cal U}  \rangle$ to $L_{\Xi}$ in Eq.~(\ref{effLag}). 
$V^{\cal U} \equiv {\cal U} V {\cal U}^{-1}$
is given by Eq.~(\ref{VCoul}) or Eq.~(\ref{Vdual}) with 
each charge $R^{jk}_{\bf p}$ replaced by the rotated charge  
$(R^{jk}_{\bf p})^{\cal U} \equiv e^{i \Xi R} R^{jk}_{\bf p} e^{-i \Xi R}$;
the normal-ordered nature of the products is thereby left intact.  
Thus, e.g., 
\begin{equation}
\langle V^{\cal U} \rangle 
= {1\over{2}} \sum_{\bf p}
v_{\bf p} \gamma_{\bf p}^{2}\, g^{jk}_{\bf p}\, g^{mn}_{\bf -p}\, 
\langle\,  :\! (R^{jk}_{\bf -p})^{\cal U} \, (R^{mn}_{\bf p})^{\cal U}\!:\, \rangle.
\label{VUnormal_order}
\end{equation}
The crossing relation~(\ref{Xrel}) is also promoted to the dressed form,
\begin{equation}
: \!(R^{jk}_{\bf -p})^{\cal U} (R^{mn}_{\bf p})^{\cal U}\!: \ 
= - {1\over{\bar{\rho}}}  \sum_{\bf k}  
e^{i\, \ell^{2}  {\bf p \times k}}
 :\! (R^{mk}_{\bf -k})^{\cal U} (R^{jn}_{\bf k})^{\cal U}\!:  .
\label{Xrel_dressed}
\end{equation}

The $O(\Xi^r)$ piece of  $(R_{\bf -p}^{jk})^{\cal U}$
contains an operator of the form 
$R_{\cdots; { {\bf -p}- {\bf k}_{1} - \dots - {\bf k}_{r}}}^{\cdots}$ 
under integrals over momenta ${\bf k}_{i}$ of  $r$ powers of 
$\Xi_{ {\bf k}_{i}}$. 
Accordingly, 
the $O(\Xi^{r+s})$ terms of the structure factors
$\langle {: \!(R_{\bf -p})^{\cal U}\,  (R_{\bf p})^{\cal U}\! :} \rangle$
are built, under integrals over $({\bf k}_{1}, \cdots, {\bf k}_{r+s})$, 
on factors of the form
\begin{equation}
\langle\,  :R_{ {\bf -p} - {\bf k}_{1} - \dots - {\bf k}_{r} } 
R_{ {\bf p}+ {\bf k}_{r+1} + \cdots + {\bf k}_{r+s} }: \, \rangle.
\end{equation}
Associated with these are the delta-function pieces
$\propto \delta_{ {\bf p} + {\bf k}_{1} + {\bf k}_{2} + \dots +{\bf k}_{r}, {\bf 0}}$.
The remaining pieces are not $\lq\lq$constants" any more 
and now consist of products of exponentials (sines and cosines) in ${\bf p}$, 
such as 
$e^{i\ell^2 {\bf p} \times {\bf k}_{1}}$ and 
$e^{i\ell^2 {\bf k}_{2} \times ({\bf p} +{\bf k}_{1})}$,
as seen from Eq.~(\ref{RupU});
that is, they are periodic functions in ${\bf p}$ and 
their Fourier images $({\bf p \rightarrow k} )$ 
are a variety of monochromatic spectra, i.e., delta functions 
in dual variable ${\bf k}$. 
Thus the (nonsingular) oscillating pieces are again expressed 
in terms of the (singular) delta-function pieces in the dual expression.
The relevant singular pieces 
$\propto \delta_{ {\bf p}+ {\bf k}_{1} + {\bf k}_{2} + \dots+{\bf k}_{r}, {\bf 0}}$ 
are uniquely summarized by 
the expectation value  $\langle (R_{\bf -p})^{\cal U} \rangle$. 
This means that 
$\langle {: \!(R_{\bf -p})^{\cal U}\,  (R_{\bf p})^{\cal U}\! :} \rangle$ 
is calculable via  $\langle (R_{\bf p})^{\cal U} \rangle$, i.e., 
Eq.~(\ref{RRfromR})  is also promoted to rotated charges,
\begin{equation}
\big\langle {:\! (R^{jk}_{\bf -p})^{\cal U} (R^{mn}_{\bf p})^{\cal U}\! :} \big\rangle
= Z^{jk|mn}_{\bf p}
-{1\over{\bar{\rho} }}\sum_{\bf k}  e^{ i \ell^{2}  {\bf p \times k}}\, Z^{mk| jn}_{\bf k} 
\label{keyFormula}
\end{equation}
with $Z^{jk | mn}_{\bf p} \equiv
\langle (R^{jk}_{\bf -p})^{\cal U}\rangle \langle (R^{mn}_{\bf p})^{\cal U}\rangle$.
One can also write
 \begin{eqnarray}
\langle V^{\cal U} \rangle 
&=& {1\over{2}} \sum_{\bf p}
v_{\bf p} \gamma_{\bf p}^{2}\,  g^{jk; a}_{\bf p} g^{mn;b}_{\bf -p} 
\langle (R^{jk;aa}_{\bf -p})^{\cal U}\rangle\, 
\langle (R^{mn;bb}_{\bf p})^{\cal U} \rangle 
\nonumber\\
&&\!\!\!\! -{1\over{2 \bar{\rho}}}  \sum_{\bf k}  W^{jk;mn;ab}_{\bf k} 
\langle (R^{m k;ba}_{\bf - k})^{\cal U} \rangle 
\langle (R^{j n;ab}_{\bf k})^{\cal U} \rangle .
\label{VUexp}
\end{eqnarray}

Equation~(\ref{keyFormula}) is one of the key results of the present paper, 
and allows one to evaluate the structure factors 
$\langle\, :\! R^{\, \cal U}\, R^{\,\cal U}\! :\, \rangle$
by a far simpler calculation handling only expectation values 
$\langle R^{\, \cal U} \rangle$.
Normally great labor is needed to calculate such structure factors 
of rotated charges $R^{\cal \, U}$.
They, when expanded in powers of $\Xi$, proliferate rapidly 
in number and variety of terms,
but many of them turn out to vanish on substituting the zeroth-order factors
$\langle {:\!R R\!:} \rangle$. 
Note that in Eq.~(\ref{keyFormula}) 
integration over $\{ {\bf k}_{i} \}$ of $\{ \Xi_{{\bf k}_{i}}\}$, 
normally made toward the end of calculation, is carried out first 
in evaluating the expectation values $\langle R^{\cal \, U} \rangle$.
Our formula~(\ref{keyFormula}) or (\ref{VUexp})  
thus neatly rearranges steps of calculations 
and achieves a most efficient approach to the goal.

\section{ Coulombic corrections}

The calculation of many-body corrections $\langle V^{\cal U} \rangle$ 
is greatly simplified by use of Eq.~(\ref{VUexp}). 
See Appendix A for details.  
The $O(\Xi)$ term vanishes, and the $O(\Xi^{2})$ term is
\begin{eqnarray}
\langle V^{\cal U} \rangle &=& \bar{\rho}\,
\sum_{n> j} \sum_{b,a} \sum_{\bf k} V_{\xi}^{nj},  
\nonumber\\
V_{\xi}^{nj} &=& D^{nj;ba}_{\bf k}\, 
(\xi^{nj;bb}_{\bf k})^{\dag}\, \xi^{nj;aa}_{\bf k} \!
+ E^{nj;ba}_{\bf k} \, (\xi^{nj;ba}_{\bf k})^{\dag}\xi^{nj;ba}_{\bf k},\ \ \ \ \ 
\label{VU_intraLL}
\end{eqnarray}
with
\begin{eqnarray}
 E^{nj;ba}_{\bf k} &=& -\sum_{\bf p} v_{\bf p}\gamma_{\bf p}^{2}
\Big[ \sum_{r}\{
\nu_{r}^{b}\,|g^{n r;b}_{\bf p}|^{2} -\nu^{a}_{r}\, |g^{jr;a}_{\bf p}|^{2} \}
+ (\nu_{j}^{a}-  \nu_{n}^{b} )\, 
g^{nn;b}_{\bf -p}\, g^{jj;a}_{\bf p}\, c({\bf p,k})  \Big],
\nonumber\\
D^{nj;ba}_{\bf k} &=& {\cal N}^{aa}_{nj}{\cal N}^{bb}_{nj}\,  
 (\bar{\rho}\, v_{\bf k}\gamma_{\bf k}^{2}\,  g^{nj;b}_{\bf k}  g^{jn;a}_{\bf -k})
 \propto O(|{\bf k}|^{|n-j|}), 
\label{DkEk}
\end{eqnarray}
where $c({\bf p,k}) \equiv  \cos (\ell^{2}  {\bf p\! \times\! k})$.
The first term $\propto D^{nj;ba}_{\bf k}$ comes from the direct interaction 
and is short-ranged as it vanishes for ${\bf k}\rightarrow0$.
The exchange correction $\propto E^{nj;ba}_{\bf k}$ is composed 
of the self-energy corrections due to the filled levels
and an attraction between the excited electron and  hole pair.
The effective Lagrangian is now given by 
$L_{\Xi} = L_{\Xi}^{\rm1b} - \langle V^{\cal U} \rangle$.

Possible contributions from nontrivial intralevel correlations 
$\delta\hat{s}^{a}_{j}({\bf p})$
are readily extracted from $\langle V^{\cal U}  \rangle$ in Eq.~(\ref{VUnormal_order}).
In particular, when the initial level  $(j,a)$ is partially filled,
one can simply retain terms involving 
$\langle\, :\!R^{jj} R^{jj}\!\!:\, \rangle 
\propto \bar{\rho}\nu_{j} \delta \hat{s}_{j}$.
The result is an addition to $E^{nj;ba}_{\bf k}$ of the form 
\begin{eqnarray}
 \delta E^{nj}_{\bf k} &=& \sum_{\bf p}v_{\bf p} \gamma_{\bf p}^{2} 
 \Big[   \delta^{ba}\, \delta\hat{s}^{a}_{j}({\bf p+k}) \, |g^{nj;a}_{\bf -p}|^2  
 \nonumber\\
 &&
 +  \delta\hat{s}^{a}_{j}({\bf p}) 
\{ 
c({\bf p,k})\,g^{nn;b}_{\bf -p} g^{jj;a}_{\bf p} - |g^{jj;a}_{\bf p}|^2 \} 
\Big].
\label{deltaEk}
\end{eqnarray}
Equations~(\ref{VU_intraLL}) - (\ref{deltaEk}) agree with 
and partly generalize some earlier calculations.$^{10-12,31}$

So far we have retained valley labels for practical applications 
as well as to show the generality of  our method of calculation. 
In general, when one considers a specific $n \leftarrow j$  interlevel excitation 
one also has to take into account all such related 
excitation channels that are strongly mixed with it 
via the short-ranged direct interaction $\propto D^{nj;ba}_{\bf k}$
and eventually go through a matrix diagonalization, 
as done in the literature.$^{10,11}$

Instead of handling such a general case, we from now on focus on cyclotron resonance,
i.e., optical interlevel excitations 
at zero momentum transfer ${\bf k}\rightarrow 0$, 
where no mixing takes place, with the selection rule$^{9}$
$\Delta |n| =\pm 1$, 
i.e., (i) $n+1 \leftarrow \pm n$ and (ii)  $\pm n \leftarrow\! -(n+1)$
for $n=0,1,2, \cdots$.
The $n\leftarrow j$ resonance energy is then simply  written as 
\begin{eqnarray}
\epsilon_{\rm exc}^{n\leftarrow j} &=& \epsilon_{n}- \epsilon_{j} +\Delta \epsilon^{n,j}, 
\nonumber \\
\Delta \epsilon^{n,j}&=& -\sum_{\bf p} v_{\bf p}\gamma_{\bf p}^{2}
\Big[  \sum_{r}  \nu_{r}\,\{ |g^{n r}_{\bf p}|^{2} - |g^{jr}_{\bf p}|^{2} \} 
+ (\nu_{j}-  \nu_{n} )\, g^{nn}_{\bf -p}\, g^{jj}_{\bf p} \Big]
\label{fullEexc} 
\end{eqnarray}
for each (valley, spin) channel.
In particular, for the $1\leftarrow 0$  resonance 
\begin{eqnarray}
\Delta \epsilon^{1,0}\!\! &=&\! -\sum_{\bf p} v_{\bf p}\gamma_{\bf p}^{2}
\Big[  \sum_{r\le -1} \{ |g^{1 r}_{\bf p}|^{2} -|g^{0r}_{\bf p}|^{2} \}
\nonumber\\
&&
+\{\nu_{0} - \delta \hat{s}_{0}({\bf p})\}\, 
(|g^{1 0}_{\bf p}|^{2} -|g^{00}_{\bf p}|^{2} + g^{00}_{\bf p}\, g^{11}_{\bf -p}) \Big], 
\end{eqnarray}
where the contribution of $\delta E^{10}_{\bf k}$ is also included.
For conventional 2d electrons, 
one only has the last term $\propto \{\nu_{0} - \delta \hat{s}_{0}\}$,
though it actually vanishes in accordance with Kohn's theorem.$^{7}$ 
It happens to vanish also for ($\Delta \epsilon^{1,0}$ of) graphene, 
since 
$|g^{1 0}_{\bf p}|^{2} -|g^{00}_{\bf p}|^{2} + g^{00}_{\bf p}\, g^{11}_{\bf -p}\rightarrow 0$,
as one can verify using Eq.~(\ref{example_g}).
Thus $\Delta \epsilon^{1,0}$ consists solely of the self-energy correction 
due to the filled valence band
and is actually logarithmically divergent.

Cyclotron resonance in graphene and bilayer graphene 
was studied earlier.$^{10-12}$
Here we briefly review it and present some basic formulas 
that clarify the structure of the selfenergy corrections.  
Let us first note the completeness relation$^{24}$
\begin{equation}
\sum_{k=-\infty}^{\infty}|g^{n k}_{\bf p}|^2 
= e^{{1\over{2}}\ell^2 {\bf p}^2}
=1/ \gamma_{\bf p}^{2}
\label{completeness-rel}
\end{equation}
that, in general, holds for the eigenmodes of the one-body Hamiltonian.
The infinite sum in the self-energy corrections to level $n$ 
is thereby rewritten as
\begin{equation}
\gamma_{\bf p}^{2}\!\sum_{k\le -1}|g^{n k}_{\bf p}|^2
= {\textstyle{1\over{2}} } - {\textstyle{1\over{2}} } s_{n} F_{|n|}(z) 
 -{\textstyle{1\over{2}} } \gamma_{\bf p}^{2}\, |g^{n 0}_{\bf p}|^2 ,
\label{selfenergy_sum}
\end{equation}
where 
$z= {1\over{2}}\ell^2 {\bf p}^{2}$, $s_{n} = {\rm sign}[n] \rightarrow \pm1$ 
and 
\begin{eqnarray}
F_{n}(z) &\equiv& 
\gamma_{\bf p}^{2} \sum_{k=1}^{\infty}\{ |g^{n k}_{\bf p}|^2 - |g^{n, -k}_{\bf p}|^2 \}
\ \ (n>0), \\
&\stackrel{\mu\rightarrow0}{=}& \!\!  
e^{-z}\sum_{k=1}^{\infty} \sqrt{{k\over{n}}}\, {n!\over{k!}}\, 
z^{k-n} L^{k-n}_{n-1}(z)\,  L^{k-n}_{n}(z),
\nonumber\\
F_{0}(z)  &\stackrel{\mu\rightarrow0}{=}&  0 \ .
\label{Fnz}
\end{eqnarray}
The resonance energy corrections $\Delta \epsilon^{n,j}$ then reveal 
the underlying electron-hole $(eh)$ symmetry, 
\begin{eqnarray}
\Delta \epsilon^{n,j}&=& 
\sum_{\bf p} v_{\bf p} \left[{\textstyle{1\over{2}} } \{s_{n}F_{|n|}(z) - s_{j}F_{|j|}(z) \} 
- \gamma_{\bf p}^{2}\,  G^{nj}_{\bf p}\right],
\nonumber\\
G^{nj}_{\bf p}&=& 
\sum_{k} \nu [k]\,  ( |g^{n k}_{\bf p}|^{2} - |g^{jk}_{\bf p}|^{2}) 
+ (\nu_{j}-  \nu_{n} )\, g^{nn}_{\bf -p}\, g^{jj}_{\bf p},
\nonumber\\
\nu[k] &\equiv& 
\nu_{k}\,  \theta_{(k\ge1)} - (1-\nu_{k})\, \theta_{ (k\le -1)} 
+ (\nu_{0}- {\textstyle{1\over{2}} })\,  \delta^{k0}, 
\label{deltaEnj}
\end{eqnarray}
where  $\theta_{(k\ge1)} =1$ for $k\ge1$ and $\theta_{(k\ge1)} =0$ otherwise; 
analogously for $\theta_{(k\le -1)}$.
Here $G^{nj}_{\bf p}$ summarize corrections due to a finite number of 
electron or hole levels around the $n=0$ level,
as seen from the definition of the $eh$-symmetric filling factor $\nu[k]$.
The filled valence band also gives rise to 
$eh$-symmetric corrections $\propto s_{n} F_{|n|}(z)$.
Formulas~(\ref{completeness-rel})~-~(\ref{deltaEnj}) 
serve to provide compact analytic expressions 
for some numerically-handled portions of an earlier analysis.$^{12}$ 
They are equally  generalized to the case of few-layer graphene.

For simplicity, let us set  tiny valley breaking $\mu\rightarrow0$ below; 
actually, $\mu\not=0$ requires separate renormalization.\cite{KS_CR}
One then finds that, for $n\ge 1$,
$F_{n}(z) >0$, $F_{n}(0)=1$ and 
$F_{n}(z) \approx \sqrt{n/2}/({\ell |{\bf p}|})$
for  ${\bf p}\rightarrow\infty$, 
which reveals that the divergence in $\Delta \epsilon^{n,j}$ is 
proportional to the large-${\bf p}$ behavior of  
$s_{n} F_{|n|}(z) - s_{j}F_{|j|}(z)$, i.e., of the form 
$\propto (s_{n} \sqrt{|n|} -s_{j} \sqrt{|j|})\, \log (\ell\,  \Lambda)$, 
with momentum cutoff $\Lambda$.

This implies that the divergences in all $\Delta \epsilon^{n,j}$ 
are removed via renormalization$^{35}$ of the velocity 
\begin{equation}
v = Z_{v}\, v^{\rm ren}= v^{\rm ren} + \delta v,  
\end{equation}
with $\delta v =(Z_{v}-1)\, v^{\rm ren}$.
Indeed, setting 
$\epsilon_{n}= \epsilon_{n}^{\rm ren} + \delta_{\rm ct} \epsilon_{n}$
with renormalized energy 
$\epsilon_{n}^{\rm ren}=s_{n} \sqrt{2|n|}\, v^{\rm ren}/\ell$ 
and the counterterm 
$\delta_{\rm ct} \epsilon_{n} \propto s_{n}\sqrt{|n|}\, \delta v$,
one can rewrite the excitation energy as
\begin{equation}
\epsilon_{\rm exc}^{n \leftarrow j} 
= \epsilon_{n}^{\rm ren} -\epsilon_{j}^{\rm ren} 
+ (\triangle \epsilon^{n, j})^{\rm ren}.
\end{equation}
All the corrections $(\triangle \epsilon^{n,j})^{\rm ren} 
\equiv \delta_{\rm ct}\epsilon_{n}- \delta_{\rm ct}\epsilon_{j} 
+ \triangle \epsilon^{n,j}$ 
are now clearly made finite by a single choice of  $\delta v$.

Let us choose $\delta v$ so that $ (\triangle \epsilon^{1, 0})^{\rm ren}=0$, or 
\begin{equation}
\delta v = - \textstyle {\ell\over{\sqrt{2}}}\, \Delta \epsilon^{1,0}
= - {1\over{8}}\, (\alpha/\epsilon_{b})\, [\log (\Lambda^2\ell^2) + {\rm const.} ] ;
\end{equation}
this defines $v^{\rm ren}$ 
via  $\epsilon_{\rm  exc}^{1\leftarrow 0} =\omega_{c}^{\rm ren}
\equiv \sqrt{2}\, v^{\rm ren}/\ell$
at each value of $B$.
The renormalized corrections $(\triangle \epsilon^{n, j})^{\rm ren}$ 
are thereby given by $\triangle \epsilon^{n, j}$ in Eq.~(\ref{deltaEnj}) 
with $F_{|n|}(z)$ and $F_{|j|}(z)$ replaced by the renormalized counterparts
\begin{equation}
F_{|m|}^{\rm ren}(z)= F_{|m|}(z) 
- \sqrt{|m|}\, \{F_{1}(z) -\gamma_{\bf p}^2\,  g^{11}_{\bf -p}\, g^{00}_{\bf p}\}.
\end{equation}
A key effect of renormalization is the fact that,
as implied by 
$v=v^{\rm ren}|_{B} + \delta v|_{B}$, 
the renormalized velocity runs with the magnetic field,
\begin{equation}
v^{\rm ren}|_{B} = v^{\rm ren}|_{B_{0}}
- {\alpha\over{8\epsilon_{b}}}\, \log (B/B_{0}).
\end{equation}

\begin{figure}[bt]
\centerline{\psfig{file=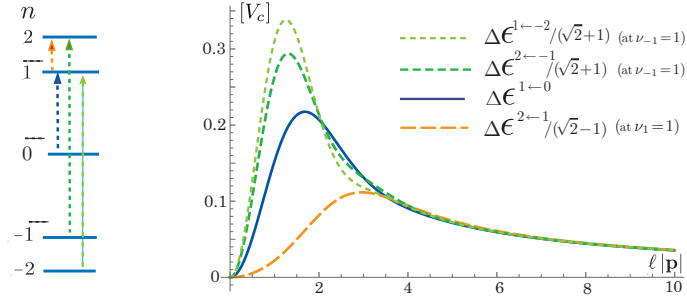,width=3.55in}}
\vspace*{8pt}
\caption{Momentum profiles of (rescaled) many-body corrections
$\triangle \epsilon^{n, j}/ (s_{n}\sqrt{|n|} -s_{j}\sqrt{|j|})$ for some typical channels.
}
\end{figure}


Figure~2 shows momentum profiles of some rescaled corrections
$\triangle \epsilon^{n, j}/ (s_{n}\sqrt{|n|} -s_{j}\sqrt{|j|})$
associated with  levels $n=0,\pm 1, \pm2$.
Changes in profile relative to the $1 \leftarrow 0$ profile represent 
the genuine corrections after renormalization. 
The general tendency is that 
such renormalized corrections are  negative for intraband resonance 
and positive for interband resonance. 
Actually this is consistent with an experiment$^{14}$ 
which observed an appreciable deviation 
of the ratio of $\epsilon_{\rm exc}^{1 \leftarrow 0}$ 
and $\epsilon_{\rm exc}^{2 \leftarrow -1}$ 
from the tree-level value $1: (1+\sqrt{2})$. 
Running of the renormalized velocity has also been observed$^{16}$
in experiment.

\section{Intralevel excitations}

In this section we examine some examples of collective excitations 
within a Landau level.
The first example is a monolayer prototype with ordinary electrons 
of spin up and down 
occupying the $n=0$ Landau level at half filling $\nu=1$.  
Let us form the spin doublet 
$\psi =(\psi^{0}_{\uparrow}, \psi^{0}_{\downarrow} )^{\rm t}$ 
and denote the spin charge as 
$\rho^{\mu} =\psi^{\dag} {1\over{2}} (\sigma^{0}, \sigma^{a})\, \psi$
with $\sigma^{0}=1$ and Pauli matrices $\{\sigma^{a}\}$;
we denote the Fourier image as 
$\rho^{\mu}_{\bf -p} = \gamma_{\bf p}\, R^{\mu}_{\bf -p}$ and 
$R^{\mu}_{\bf -p} =\int d y_{0}\, 
\psi^{\dag} {1\over{2}} \sigma^{\mu}e^{i {\bf p\cdot r}} \psi$.
The ground state $|{\rm Gr}\rangle$ of such a system 
is characterized by density 
$\langle \rho^{0}\rangle ={1\over{2}}\, \bar{\rho}$ 
and spin $\langle \rho^{a}\rangle ={1\over{2}}\, \bar{\rho}\, n^{a}$ 
with ${\bf n} \equiv (n^{1}, n^{2}, n^{3})$ pointing in a definite direction 
in spin space; ${\bf n} \cdot {\bf n}=1$.

The Zeeman energy $- \mu_{\rm Z}\, \rho^{3}$ naturally favors $n^{3}=1$
whereas, in the $\mu_{\rm Z} \rightarrow 0$ limit, 
spin ${\bf n}$ may take any direction,
giving rise to spontaneous spin coherence.\cite{Moon}
Such a ground state supports collective spin excitations via Coulomb interactions.
Let us now study their spectra in some detail.

As before, we describe such spin waves 
as a local spin rotation ${\cal U}^{-1} |{\rm Gr}\rangle$ of $|{\rm Gr}\rangle$
with ${\cal U} = e^{i \Xi R}$, where $\Xi R$ now stands for 
$\sum_{\bf p} \sum_{a=1}^{3}\Xi^{a}_{\bf p} R^{a}_{\bf -p}$.
 We thus consider $L_{\Xi}$ in  Eq.~(\ref{effLag}) and 
now substitute for $H^{\rm tot}$
the Coulomb interaction acting within the $n=0$ level  
\begin{equation}
V_{0}= 2\sum_{\bf p}v_{\bf p}\gamma_{\bf p}^{2}\, 
:R^{0}_{\bf -p}R^{0}_{\bf p}: .
\label{Vzeroth}
\end{equation}
For $R^{\mu}_{\bf p}$ the crossing relation~(\ref{Xrel}) takes the form 
\begin{equation}
: R^{\alpha}_{\bf -p} R^{\beta}_{\bf p}\!: \,  
 = -{1\over{\bar{\rho}}}\sum_{\bf k}  
e^{i\, \ell^{2}  {\bf p \times k}}\,  
 : R^{\mu}_{\bf -k} R^{\nu}_{\bf k}:\, T^{\mu\nu|\alpha\beta},
\end{equation}
where $T^{\mu\nu|\alpha\beta}
= {1\over{4}} {\rm tr}(\sigma^{\alpha}\sigma^{\mu} \sigma^{\beta} \sigma^{\nu} )$ 
and Greek letters run over $(0,1,2,3)$.
$V_{0}$ is thereby cast into the dual form
\begin{equation}
V_{0} 
= -  (1/\bar{\rho}) \sum_{\bf k} v^{\rm dual}_{\bf k}
:R^{a}_{\bf -k} R^{a}_{\bf k} +R^{0}_{\bf -k} R^{0}_{\bf k}:
\label{Vzerodual}
\end{equation}
with a sum over $a \in (1,2,3)$ and 
\begin{eqnarray}
 v^{\rm dual}_{\bf k} &\equiv & \sum_{\bf p} v_{\bf p}\gamma_{\bf p}^{2}
e^{i\, \ell^{2}  {\bf p \times k}}
=V_{c}\,\textstyle \sqrt{{\pi\over{2}}}\,I_{0}(z)\, e^{-z}, \nonumber\\
&=&  V_{c}\,\textstyle \sqrt{{\pi\over{2}}}\, \{1 -  {1\over{4}}\, \ell^2{\bf k}^2 
+ {3\over{64}}\, (\ell\, |{\bf k}|)^4 + \cdots \},
\end{eqnarray}
where $z= {1\over{4}}\, \ell^{2}{\bf k}^{2}$ and 
$V_{c}\equiv \alpha/(\epsilon_{b} \ell)$.

Note now that the structure factors 
$\langle {:\!R^{\mu}_{\bf -p}R^{\nu}_{\bf p}\! :} \rangle$ consist of 
a $\delta_{\bf p,0}$ piece and a constant piece.
This is clear for ${\bf n}=(0,0,1)$, in which case 
only $\langle\,  {:\!R^{0}R^{0}\!:}\, \rangle$  and
$\langle\, {:\!R^{3}R^{3}\!:}\, \rangle$ are nonzero.
For general ${\bf n}$, 
the associated structure factors are constructed 
from  this $\langle\, {:\!R^{3}R^{3}\!:} \, \rangle$
by a global rotation  in spin space, and naturally share the structure
$\propto  \delta_{\bf p,0}$ piece + constant.
As a result,   
analogues of Eqs.~$(\ref{RRfromR})$ and~$(\ref{keyFormula})$
again hold for $R^{\mu}_{\bf p}$,
and one can calculate
$\langle\, {:\! (R^{\mu}_{\bf p})^{\cal U}\,  (R^{\nu}_{\bf -p})^{\cal U}\! :} \, \rangle$ 
via the expectation values 
$\langle (R^{\mu}_{\bf p})^{\cal U} \rangle$ alone.
The latter read,  to $O(\Xi^{2})$, 
\begin{eqnarray}
\langle (R^{a}_{\bf -p})^{\cal U} \rangle 
&=&  {1\over{2}}\,  \bar{\rho}\,  \big\{ n^{a} \delta_{\bf p,0} 
+  \tilde{\Xi}_{\bf -p}^{ab} n^{b} 
+ {1\over{2}} \sum_{\bf k} c_{\bf k,p}\,  
 \tilde{\Xi}_{\bf k}^{ac} \tilde{\Xi}_{\bf -p-k}^{cb} n^{b} \big\},
 \nonumber\\
\langle (R^{0}_{\bf -p})^{\cal U}\rangle 
&=& {1\over{2}}\,  \bar{\rho}\, \big\{ n^{0} \delta_{\bf p,0} 
+ {1\over{2}}  \sum_{\bf k} s_{\bf k,p}\, 
 \Xi_{\bf k}^{a}\, \tilde{\Xi}_{\bf -k-p}^{ab}\, n^{b} \big\}, 
\end{eqnarray}
where 
$c_{\bf k,p}\equiv \cos ({1\over{2}} \ell^2 {\bf k\!\times\! p})$
and $s_{\bf k,p}\equiv \sin ({1\over{2}} \ell^2 {\bf k\!\times\! p})$;
$\tilde{\Xi}_{\bf k}^{ac} \equiv \epsilon^{abc}\, \Xi_{\bf k}^{b}$
and $\epsilon^{abc}$ is the totally-antisymmetric tensor 
with $\epsilon^{123}=1$. 
Note that for ${\bf p}\rightarrow 0$  
the 
$\sum_{\bf k} s_{\bf k,p}\, 
\Xi_{\bf k}^{a}\, \tilde{\Xi}_{\bf -k-p}^{ab}\, n^{b}$ term
in $\langle (R^{0}_{\bf -p})^{\cal U}\rangle$ is reduced 
to a surface integral equal to $4\pi\ell^2\,   Q_{\rm top}$, 
where 
$Q_{\rm top} = (8\pi)^{-1} \int d^{2}{\bf x}\, 
\epsilon_{ij}\, \epsilon^{abc}\, \Xi^{a} (\partial_{i}\Xi^{b})\, (\partial_{j} \Xi^{c})$ 
is the topological charge$^{32}$
carried by the spin wave $\Xi^{a}$.

Noting Eqs.~(\ref{Vzeroth}) and~(\ref{Vzerodual}),
one can calculate 
$V_{\rm eff} \equiv \langle {\cal U} V_{0}\, {\cal U}^{-1} \rangle$
via
\begin{equation}
V_{\rm eff}
= {1\over{\bar{\rho}}}  \sum_{\bf p} 
\Big[ (2\bar{\rho}\, v_{\bf p}\gamma_{\bf p}^{2} -  v^{\rm dual}_{\bf p})\, 
Z^{00}_{\bf p} -  v^{\rm dual}_{\bf p}\,  Z^{aa}_{\bf p}\Big] ,
\end{equation}
with $Z^{\mu\nu}_{\bf p} \equiv
\langle (R^{\mu}_{\bf -p})^{\cal U}\rangle \langle (R^{\nu}_{\bf p})^{\cal U}\rangle$.
The result to $O(\Xi^{2})$ is
\begin{equation}
V_{\rm eff} = -{\textstyle{1\over{2}}} \, v^{\rm dual}_{\bf p\rightarrow 0}\,
\Big( \int \! d^{2}{\bf x}\,  \bar{\rho}  +  Q_{\rm top} \Big) 
+{\textstyle{1\over{4}}}\, \bar{\rho} \sum_{\bf p} 
(v^{\rm dual}_{\bf p=0} -v^{\rm dual}_{\bf p})\, 
(\tilde{\Xi}_{\bf -p}^{ac}n^{c} ) \,   ( \tilde{\Xi}_{\bf p}^{ab}\,  n^{b}). \ \ 
\end{equation}

On the other hand, the ${\cal U} i \partial_{t}  {\cal U}^{-1}$ term leads to 
$L_{t} = - {1\over{4}} \bar{\rho} \sum_{\bf p} \epsilon^{abc} 
n^{a}\, \Xi^{b}_{\bf -p}\, \dot{\Xi}^{c}_{\bf p}$,
and the Zeeman energy 
$H_{Z} = -\mu_{Z}\,\langle (R_{\bf p=0}^{3})^{\cal U} \rangle
=- {1\over{2}} \,\bar{\rho}\,  \mu_{Z}\, \int d^{2}{\bf x}\, n^{3} + \cdots$
selects the spin direction $n^{3}=1$.
One can now write down the effective Lagrangian 
$L_{\rm eff} \sim L_{t} - V_{\rm eff} - H_{Z}$ for $\Xi^{a}$,
\begin{equation}
L_{\rm eff} 
=  {\textstyle {1\over{4}} }\, \bar{\rho} \sum_{\bf p}
\phi^{\dag}_{\bf p}\, \big[ i\partial_{t} - (\epsilon_{\bf p} + \mu_{Z}) \big]\, \phi_{\bf p},
\label{LeffSpinWave}
\end{equation}
where 
$\phi_{\bf p} = \Xi^{1}_{\bf p} + i\Xi^{2}_{\bf p}$
and $\phi^{\dag}_{\bf p} = \Xi^{1}_{\bf -p} - i\Xi^{2}_{\bf -p}$.
The excitation spectrum consists of the Zeeman gap $ \mu_{Z}$ and
the Coulombic energy
\begin{equation}
\epsilon_{\bf p}=v^{\rm dual}_{\bf p=0} -v^{\rm dual}_{\bf p} 
= \textstyle \sqrt{{\pi\over{2}}}\, V_{c}\, \{{1\over{4}}\, \ell^2{\bf p}^2 + \cdots\},
\end{equation}
in agreement with an earlier result\cite{KH};
this $\epsilon_{\bf p}$ rapidly rises 
with increasing $|{\bf p}|$ and approaches 
$v^{\rm dual}_{\bf 0}=\sqrt{\pi/2}\, V_{c}$
for ${\bf p}\rightarrow\infty$.
In the limit $\mu_{Z}\rightarrow 0$, the spin ${\bf n}$ can point in any direction 
and spin waves have the spectrum $\epsilon_{\bf p}$.

This Lagrangian $L_{\rm eff}$ applies to some other cases as well.  
(i) Valley pseudospin waves in graphene. 
Consider, e.g., the $\nu=0$ vacuum state in graphene 
with degenerate valley $(\delta m\rightarrow 0)$ 
and frozen spin (via the Zeeman energy). 
The half-filled $n=0$ level will then support a local valley excitation
which is described by this  $L_{\rm eff}$ 
with $\mu_{Z}\rightarrow 0$.
(ii) Layer excitations in bilayer systems.
Consider a bilayer system of conventional electrons with frozen spin, 
and replace, in the above analysis, the spin by the layer,
with  $(\uparrow, \downarrow)$ now reinterpreted as (upper layer, lower layer).
Then $L_{\rm eff}$ (with $\mu_{Z}\rightarrow 0$) 
describes local layer-pseudospin excitations
in the half-filled $n=0$ level (at  $\nu=1$) of such a bilayer system$^{32}$  
(in the limit of zero separation).

Let us here look into the latter case in more detail, 
especially to see how our present framework
streamlines actual calculations. 
In bilayer systems the difference between intralayer and interlayer Coulomb potentials
($v_{\bf p}$ and $e^{-d|{\bf p}|}\, v_{\bf p}$ with interlayer separation $d$)
gives rise to layer SU(2) breaking, 
which drives spontaneous layer coherence,$^{32}$ 
with equal population of electrons in both layers.
To verify this let us try to improve $L_{\rm eff}$ in Eq.~(\ref{LeffSpinWave}).

For a bilayer with separation $d$, 
the Coulomb interaction 
takes the form of $V_{0}$ in Eq.~(\ref{Vzeroth}) with 
$v_{\bf p}\, R^{0}_{\bf -p}R^{0}_{\bf p}$ replaced by 
$v_{\bf p}^{+}\, R^{0}_{\bf -p}R^{0}_{\bf p} 
+ v_{\bf p}^{-}\, R^{3}_{\bf -p}R^{3}_{\bf p}$,
where  $ v^{\pm}_{\bf p}= {1\over{2}}(1\pm e^{-d|{\bf p}|})\, v_{\bf p}$.
Thus the addition to $V_{0}$ is of the form
\begin{equation}
\Delta V=
2\,  \sum_{\bf p}v_{\bf p}^{-}\,\gamma_{\bf p}^{2}
:\! (R^{3}_{\bf -p}R^{3}_{\bf p} -R^{0}_{\bf -p}R^{0}_{\bf p}) \!: .
\end{equation}
Since $v^{-}_{\bf p} >0$, $\Delta V$ favors layer pseudospin $n^{3}=0$,
 i.e., equal population in both layers.
Its dual form is 
\begin{equation}
\Delta V = -{2\over{\bar{\rho}}} \sum_{\bf k}
v^{- {\rm dual}}_{\bf k} :\! (R^{3}_{\bf -k}R^{3}_{\bf k} 
-R^{a}_{\bf -k}R^{a}_{\bf k} )\! :\ ,
\end{equation}
with 
$v^{- \,{\rm dual}}_{\bf k} \equiv \sum_{\bf p} v^{-}_{\bf p}\gamma_{\bf p}^{2}\, 
e^{i\, \ell^{2}  {\bf p \times k}}$, or
\begin{equation}
v_{\bf p}^{- {\rm dual}}= \textstyle {1\over{2}} V_{c} \big\{ 
\hat{d} -{1\over{2}}\sqrt{\pi\over{2}}\, \hat{d}^{2} 
-({1\over{2}}\, \hat{d}  - {3\over{8}}\sqrt{{\pi\over{2}}} \, \hat{d}^{2} )\,q^2+ \cdots  
\big\},
\end{equation}
where 
$\hat{d} = d/\ell$ and $q= \ell |{\bf p}|$.

It is now a simple task to calculate the layer breaking correction
$\Delta V_{\rm eff} = \langle (\Delta V)^{\cal U} \rangle$
via the formula 
\begin{equation}
\Delta V_{\rm eff} =
{2\over{\bar{\rho}}} \sum_{\bf p} 
\big[ -\bar{\rho}\, v^{-}_{\bf p}\gamma_{\bf p}^{2}\, Z^{00}_{\bf p} 
+ v^{- {\rm dual}}_{\bf p}\, Z^{aa}_{\bf p} 
+ (\bar{\rho}\, v^{-}_{\bf p}\gamma_{\bf p}^{2}
- v^{- {\rm dual}}_{\bf p})  Z^{33}_{\bf p} \big].
\end{equation}
The result is
\begin{eqnarray}
\Delta V_{\rm eff} &=&
{\textstyle{1\over{4}}} \bar{\rho} \sum_{\bf p}\, 
\big\{ 
 \beta^{(2)}_{\bf p}\, \hat{\Xi}^{2}_{\bf -p} \hat{\Xi}^{2}_{\bf p}
+  \beta^{(3)}_{\bf p}\, \Xi^{3}_{\bf -p}\Xi^{3}_{\bf p} 
\big\} , \nonumber\\
\beta^{(2)}_{\bf p} &\equiv& 
2\,(\rho_{0} v_{\bf p}^{-} \gamma_{\bf p}^{2} -v^{-\, {\rm dual}}_{\bf p = 0} )
= \textstyle V_{c}\, \{ {1\over{2}}\sqrt{\pi\over{2}}\, \hat{d}^{2} 
- {1\over{2}}  \hat{d}^2\, q - {1\over{2}}\, \hat{d}\, q^2 + \cdots \},
\nonumber\\
\beta^{(3)}_{\bf p} 
&\equiv& 2\, (v^{-\, \rm dual}_{\bf p}  -v^{-\, \rm dual}_{\bf p=0}) 
=- V_{c}\, \textstyle \{{1\over{2}}\, \hat{d}\, q^2 + \cdots\},
\nonumber\\
\hat{\Xi}^{2}_{\bf p} 
&\equiv&  n^{1}\, \Xi^{2}_{\bf p}- n^{2}\, \Xi^{1}_{\bf p} ,
\end{eqnarray}
where only $O(\Xi^{2})$ terms of our concern are shown.

The full effective Lagrangian is again cast in the form of 
$L_{\rm eff}$ in Eq.~(\ref{LeffSpinWave}) with the spectrum and field,
\begin{eqnarray}
\epsilon^{\rm exc}_{\bf p} &=& 
\sqrt{ ( \epsilon_{\bf p} +\beta^{(2)}_{\bf p})
( \epsilon_{\bf p} +\beta^{(3)}_{\bf p}) },  \nonumber\\
\phi_{\bf p} &=& \alpha_{\bf p}\,  \hat{\Xi}^{2}_{\bf p} 
+ i(1/\alpha_{\bf p})\, \Xi^{3}_{\bf p},
\end{eqnarray}
where 
$\alpha_{\bf p} = [( \epsilon_{\bf p} +\beta^{(2)}_{\bf p})/
( \epsilon_{\bf p} +\beta^{(3)}_{\bf p})]^{1/4}$. 
Note, in particular, that the presence of the $O(|{\bf p}|^{0})$ term 
in $\beta^{(2)}_{\bf p}$ critically changes the long-wavelength property 
of the pseudospin wave, 
i.e., from  $\epsilon_{\bf p} \sim O({\bf p}^2)$ to
\begin{equation}
\epsilon^{\rm exc}_{\bf p} \approx  |{\bf p}|\, V_{c}\sqrt{\pi}\,
{\textstyle{1\over{4}} }\, d\, (1-\sqrt{8/\pi}\, d/\ell\, )^{1/2}.
\end{equation}
These results reproduce and partly generalize 
those of earlier studies.$^{32}$

\section{Intralevel density  excitations}

So far we have handled collective excitations described as rotations 
in orbital space or in spin or valley space. 
In this section we consider how to treat intralevel density excitations
such as those$^{30}$ over the FQH states. 
For definiteness let us consider (single-spin) density fluctuation 
within the partially-filled $n$th level (with $0<\nu_{n}<1$), 
described by a trial state $R^{nn}_{\bf -p}|{\rm Gr}\rangle$.
Such a density excitation is not described by a phase rotation alone 
and requires an amplitude modulation as well. 
Indeed, for a real phase $(\Xi^{nn}_{\bf p})^{\dag} = \Xi^{nn}_{\bf -p}$,
Eq.~(\ref{effLag}) fails to work 
as the resulting $\sum_{\bf p}\Xi^{nn}_{\bf p}\, i \partial_{t}\Xi^{nn}_{\bf -p}$ term
vanishes (up to a total derivative).

Let us recall that, for a trial state of the form 
$R^{nn}_{\bf -p}|{\rm Gr}\rangle$,
the original SMA excitation spectrum reads$^{30}$
\begin{equation}
\epsilon_{n}({\bf p}) =
\langle R^{nn}_{\bf p}\,  [H^{\rm tot}, R^{nn}_{\bf -p}] \rangle
/ \langle R^{nn}_{\bf p} R^{nn}_{\bf -p} \rangle, 
\end{equation}
where 
$\langle \cdots \rangle  \equiv \langle  {\rm Gr}| \cdots | {\rm Gr}\rangle$.
One can thus suppose a Lagrangian of the form 
\begin{equation}
L^{\rm eff}_{\Xi} \sim  \bar{\rho}\, \nu_{n} \int dy_{0}\, \hat{s}({\bf p}) 
(\Xi^{nn}_{\bf p})^{\dag} \{i\partial_{t} - \epsilon_{n}({\bf p}) \} \Xi^{nn}_{\bf p},
\end{equation}
with $(\Xi^{nn}_{\bf p})^{\dag} \not= \Xi^{nn}_{\bf -p}$,
i.e.,  a complex field in real space.
Actually this $L^{\rm eff}_{\Xi}$follows if one sets  
\begin{equation}
L^{\rm eff}_{\Xi} = \big\langle 
i (\Xi' R)\,  \dot{\Xi} R  
- {\textstyle{1\over{2}}} [\Xi R ,  [H^{\rm tot} , \Xi R] ]\,  \big\rangle , 
 \label{LXnn_two}
\end{equation}
with
$\Xi R = \sum_{\bf p}\Xi^{nn}_{\bf p}R^{nn}_{\bf -p}$ 
and 
$\Xi' R \equiv (\Xi R)^{\dag} =
\sum_{\bf p}(\Xi^{nn}_{\bf p})^{\dag} R^{nn}_{\bf p}$; 
$\dot{\Xi} \equiv \partial_{t} \Xi$.
We set $\Xi'^{nn}_{\bf -p} \rightarrow (\Xi^{nn}_{\bf p})^{\dag}$
in $\Xi' R$ to extract the 
$\Xi^{\dag}i\dot{\Xi}$ term properly. 
In contrast,  the ${\textstyle{1\over{2}}} [\Xi R ,  [H^{\rm tot} , \Xi R] ]$ term, 
which is essentially equivalent to
$\Xi R\,  [H^{\rm tot},  \Xi R]$, is unambiguously cast into $\Xi^{\dag} \Xi$ form
by setting $\Xi^{nn}_{\bf -p} \rightarrow (\Xi^{nn}_{\bf p})^{\dag}$.

Actually Eq.~(\ref{LXnn_two}) applies to full modes $\Xi^{mn}_{\bf p}$. 
For off-diagonal modes $\Xi^{mn}$, 
$(\Xi' R)\,  \dot{\Xi} R \rightarrow {1\over{2}}[\Xi R,  \dot{\Xi} R]$,
and this $L^{\rm eff}_{\Xi}$ agrees with $L_{\Xi}$ in Eq.~(\ref{effLag})
[to $O(\Xi^{2})$].
This in turn implies that the spectrum 
$\epsilon_{n}({\bf p})$ is still calculable  
through $\langle V^{\cal U}  \rangle$ in Eq.~(\ref{VUnormal_order}) 
by setting 
$\Xi R\rightarrow \sum_{\bf p}\Xi^{nn}_{\bf p}\, R^{nn}_{\bf -p}$ there, 
as done for $\delta E^{nj}_{\bf k}$ in Eq.~(\ref{deltaEk}).
This step simplifies actual calculation.  
The result is 
\begin{eqnarray}
\epsilon_{n}({\bf p}) &=& \sum_{\bf k}v_{\bf k}\,  |g_{\bf k}^{nn}|^2
\{1- \cos (\ell^2{\bf k\!\times\! p}) \}\, M_{\bf p,k} , \nonumber\\
M_{\bf p,k}  &=& \{ \hat{s}_{n}({\bf p+k}) - \hat{s}_{n}({\bf k}) \} /\hat{s}_{n}({\bf p}).
\end{eqnarray}
This agrees with and somewhat generalizes an earlier result.$^{30}$

\section{Summary and discussion}

In this paper we have developed a new efficient algorithm for studying 
many-body effects on 2d electrons in a magnetic field,
and verified its utility by examining inter-Landau-level excitations
in graphene and some intra-Landau-level collective excitations 
in monolayer and bilayer systems.  
The key observation is the fact that in a magnetic field the Coulomb interaction obeys 
the crossing relation in Eq.~(\ref{Xrel}), which relates via the Fourier transform the small- and large-momentum behavior of the direct and exchange interactions.
This direct/exchange duality of the Coulomb interaction, 
when adapted to the SMA-dressed interaction [in Eq.~(\ref{Xrel_dressed})], 
efficiently rearranges steps of calculations and leads to an SMA algorithm,
based on formula~(\ref{keyFormula}), 
that greatly simplifies the calculation of many-body corrections at integer filling.
For a partially-filled level further corrections arise 
via the portion [$\delta \hat{s}_{n}({\bf p})$] of the static structure factors 
that comes from nontrivial intralevel correlations, 
as we have seen, especially in Sec.~7.

The basic crossing relation in Eq.~(\ref{Xrel}) or~(\ref{Xrel_dressed}) 
takes a natural and simple form for normal-ordered products 
${:\! R_{\bf p} R_{\bf -p}\!:}$ 
rather than regular products $R_{\bf p} R_{\bf -p}$.
As noted  in the course of our discussion, 
normal ordering is maintained 
within the algebra of charge operators, 
e.g., $[ {\, :\! V\! :\,}, R_{\bf p}]= \,  {:\! [V,R_{\bf p}]\!:}$.
This suggests that normal-ordered products are the natural entity to handle
in the formulation and practice of the SMA.

The usefulness of the present algorithm will be further appreciated
when one handles some more complex systems of 2d electrons, 
such as few-layer graphene 
in which a host of new  many-body effects$^{21-25}$
come into play,  
and the case of collective excitations 
over non-uniform ground states.$^{26,27}$
Research in this direction will be reported elsewhere.

\section*{Acknowledgements}

This work was supported in part by a Grant-in-Aid for Scientific Research
from the Ministry of Education, Science, Sports and Culture of Japan 
(Grant No. 17540253).

\appendix{Coulombic corrections $\langle V^{\cal U} \rangle$ }

In this appendix we outline the calculation of  the Coulombic correction 
$\langle V^{\cal U} \rangle$ using Eq.~(\ref{VUexp}).
Note first that, on integrating factors
$g^{jk}_{\bf p}\, g^{mn}_{\bf -p}= g^{jk}_{\bf p}\, (g^{nm}_{\bf p})^{*}$
over ${\bf p}$ symmetrically, only combinations with $j-k=n-m$ survive. 
One then immediately finds that the $O(\Xi)$ term 
in $\langle V^{\cal U} \rangle$ vanishes.

As  for the $O(\Xi^2)$ terms let us begin with the direct interaction: 
Extracting the $O(\Xi) \times O(\Xi)$ piece out of 
the first term in Eq.~(\ref{VUexp})
and selecting the $j \! \rightarrow\! n\rightarrow\! j$ process
gives rise to  $D^{nj;ba}_{\bf k}$ in Eq.~(\ref{DkEk}). 
Off-diagonal processes $(j\! \rightarrow n, n' \! \rightarrow j')$ 
contribute to yet higher-order corrections in $V$.
The remaining $O(\Xi^{0})\times O(\Xi^{2})$ term involves $v_{\bf p\rightarrow0}$ 
and is removed when the neutralizing background is taken into account.

The exchange interaction has the main structure 
$I = g^{jk;a}_{\bf p}\, g^{mn;b}_{\bf -p}$ 
$\langle (R^{mk;ba}_{\bf -k})^{\cal U} \rangle 
\langle(R^{jn;ab}_{\bf k})^{\cal U}\rangle\,
e^{i\, \ell^{2}  {\bf p \times k}}$.
Of the $O(\Xi^{0}) \times O(\Xi^{2})$ portion of $I$ 
only the diagonal  $(j\rightarrow n\rightarrow j)$ process survives 
after ${\bf p}$ integration, yielding
\begin{equation}
I^{(0,2)} = 2 \delta_{\bf k,0}\, \bar{\rho}^{2} \sum_{r,n,j} 
 |g^{jr;a}_{\bf p}|^{2} \nu_{r}^{a}\,(\nu_{n}^{c}  -\nu_{j}^{a} )
 \sum_{\bf  q} \Xi^{jn;ac}_{\bf q}\, \Xi^{nj;ca}_{\bf -q}.
\end{equation}
This leads to the selfenergy term in  $E^{nj;ba}_{\bf k}$ of Eq.~(\ref{DkEk}).

 On the other hand, selecting the diagonal process 
 out of the $O(\Xi) \times O(\Xi)$ portion yields
\begin{equation}
I^{(1,1)} \approx  \bar{\rho}^{2} \sum_{n,j} g^{jj;a}_{\bf p}\, g^{nn;b}_{\bf -p}\, 
( \nu_{n}^{b} -  \nu_{j}^{a} )^{2}\, \Xi^{jn;ab}_{\bf -k} \Xi^{nj;ba}_{\bf k}\,
e^{i\, \ell^{2}  {\bf p \times k}}, 
\end{equation}
which leads to the remaining attraction term 
in $E^{nj;ba}_{\bf k}$.


\end{document}